\def\year{2021}\relax
\documentclass[letterpaper]{article} 
\usepackage{aaai21}  
\usepackage{times}  
\usepackage{helvet} 
\usepackage{courier}  
\usepackage[hyphens]{url}  
\usepackage{graphicx} 
\usepackage{natbib}  
\usepackage{caption} 
\usepackage{amsmath}
\usepackage{amsfonts}
\usepackage{amsxtra}

\DeclareMathOperator*{\argmax}{\mathrm{argmax}}

\newcommand{\mat}[1]{\mathbf{#1}}

\newcommand{\A}{\mathcal{A}}
\renewcommand{\P}{\mat{P}}
\renewcommand{\S}{\mathcal{S}}

\newcommand{\PP}[2][]{\mathbb{P}_{#1}\left[#2\right]}
\newcommand{\EE}[2][]{\mathbb{E}_{#1}\left[#2\right]}

\title{A Methodology for the Development of \\RL-Based Adaptive Traffic Signal Controllers}
\author {
    Guilherme S.\ Varela,\textsuperscript{\rm 1}
    Pedro P.\ Santos,\textsuperscript{\rm 1}
   Alberto\ Sardinha,\textsuperscript{\rm 1}
    Francisco S.\ Melo\textsuperscript{\rm 1}\\
}
\affiliations {
    \textsuperscript{\rm 1} INESC-ID and Instituto Superior T\'{e}cnico\\
University of Lisbon, Portugal\\
    guilherme.varela@tecnico.ulisboa.pt, pedro.pinto.santos@tecnico.ulisboa.pt,\\
    jose.alberto.sardinha@tecnico.ulisboa.pt,
    fmelo@inesc-id.pt
}
\begin{document}

\maketitle
\begin{abstract}
This article proposes a methodology for the development of adaptive traffic signal controllers using reinforcement learning. Our methodology addresses the lack of standardization in the literature that renders the comparison of approaches in different works meaningless, due to differences in metrics, environments and even experimental design and methodology. The proposed methodology thus comprises all the steps necessary to develop, deploy and evaluate an adaptive traffic signal controller---from simulation setup to problem formulation and experimental design. We illustrate the proposed methodology in two simple scenarios, highlighting how its different steps address limitations found in the current literature.
\end{abstract}

\section{Introduction}
\label{Sec:Intro}
Traffic congestion is a cross-continental problem. In the United States alone, an average automobile commuter spends $54$ hours in congested traffic and wastes $21$ gallons of fuel due to congestion, leading to a total estimated cost of $1,080$~USD in wasted time and fuel per commuter \cite{schrank19tr}, not considering external costs such as the increasing price of goods caused by the inflation of transportation costs, environmental and productivity impacts, as well as the decrease of population's quality of life \cite{hilbrecht14wlj}. Similarly, a recent study shows that, in the EU, the total external costs associated with traffic is over $300,000$ million euros \cite{becker16tr}. Hence, there have been numerous initiatives to mitigate traffic congestion, such as investment in public transit systems \cite{harford06trd} or intelligent transportation systems \cite{dimitrakopoulos10vtm}. 

Traffic signals, being a fundamental element in traffic control and regulation, are at the same time responsible for a significant percentage of traffic bottlenecks in urban environments, and play a key role in addressing the problem of traffic congestion. Effective traffic signal control is, therefore, a key part of urban traffic management. Classic traffic signal control approaches from transportation engineering, such as the Webster \cite{webster_1958} or Max-Pressure \cite{varaiya_2013} methods, are capable of greatly increasing the efficiency of traffic infrastructures all around the world. However, such approaches are either unable to adapt to changing traffic volumes, or rely on oversimplified traffic models, manual-tuning and inaccurate traffic information \cite{zheng2019DiagnosingReinforcementLearning,wei_survey_2020}.


Alternatively, {\em adaptive traffic signal control} (ATSC) approaches seek to take advantage of the multiple sources of information currently available (from mobile navigation applications, ride sharing platforms, etc.). Machine learning techniques can use the data made available by such platforms to provide traffic signal control strategies that adapt to the current traffic conditions in an effective manner, providing a promising alternative to classical approaches. 

Recently, several researchers have addressed ATSC using Markov decision processes (MDP) \cite{wang_enhancing_2019,wei_survey_2020}. MDPs model discrete-time stochastic control problems, and are extensively used in artificial intelligence to describe problems of sequential decision-making under uncertainty \cite{puterman05}. In an MDP, an agent (the controller) interacts sequentially with an environment by selecting actions based on its observation of the environment's state; the actions selected by the agent influence how the environment's state evolves, and the agent receives a numerical evaluation signal (a reward) that instantaneously assesses the quality of the agent's action. The agent's goal is to select its actions to maximize some form of cumulative reward. MDPs are the backbone of {\em reinforcement learning} (RL), a machine learning paradigm in which the agent learns the optimal way of selecting the actions from direct interaction with the environment, without resorting to any pre-defined model thereof \cite{Sutton98}. 

RL algorithms are a natural choice when addressing ATSC, since they can be trained directly in the data available, without requiring human annotators to define what is a ``good'' or ``bad'' control strategy. Unfortunately, the use of RL in this domain is not without its own challenges.

One of the first challenges is the lack of standard environments in the domain of ATSC. The RL field has benefited from standardized environments and easy to use APIs that allow researchers to compare different approaches to the same problem space, e.g the Deep Q-network \cite{mnih_2013} which has been shown to generate relevant features for the Arcade Learning Environment. The need for standardization has sparkled new research towards open source frameworks \cite{genders_open-source_2019}, as means to prevent researchers to re-implement the same set of fundamental tools with which to conduct \textit{de facto} experiments. We argue it is important that the community agrees on a set of benchmark environments/traffic networks that may be used as a first test stage for the algorithms explored in the context of ATSC. The existence of such benchmarks would enable a proper comparison of different models and algorithms in a common set of environments, enabling a clearer assessment of the strengths and weaknesses of different alternatives. 

Another major challenge is related to the \textit{security} and \textit{explainability} of current RL architectures. Although some RL systems have been quite successful in improving metrics of interest such as the average travel time, some protocols are not viable to be implemented in a real world situations for a variety of reasons \cite{Ault2020LearningInterpretable} such as, long and unsafe tuning process, opaque policies, and the predominant use of synthetic simulation scenarios.

One third challenge is {\em reproducibility}. Reproducibility is thoroughly documented in RL literature, as such systems might overfit to the training experience, showing good performance during training but performing poorly at deployment time \cite{whiteson_protecting_2011}. Works show that simply changing the random seeds used to generate the simulations influence in a statistically significant manner the outcome of the RL algorithm \cite{Aslani17}.


This paper contributes one further step towards a wider application of RL in ATSC. However, for this potential to be realized, it is paramount to address the standardization, security and reproducibility issues identified above. Our contribution in this article is, therefore, a methodology for the development of RL-based adaptive traffic signal controllers that ensures some level of standardization at the different stages of the experimental process: simulation setup, environment modeling, experimental design and result reporting. We adopt an action definition which is rather constrained: it produces synchronous joint action schemes across the network -- we show that the reinforcement learning agent is able to find policies that are on par with classical controllers which benefit from both human supervision and from decades-old literature from the Transportation domain. Such action plans generate protocols which are more in line with governmental transportation department's expectations, hence they provide more trust in face of the liability that such regulatory agencies face. In particular, with respect to experimental design and result reporting, we discuss good practices and relevant metrics that have been explored in different works, highlighting their merits and how their adoption may contribute to better interpretation of experimental results and mitigate reproducibility issues. We illustrate our own methodology by applying its different steps in designing a traffic signal controller using the well known deep $Q$-network (DQN) algorithm \cite{mnih_2013}.

\section{Background}
\label{Sec:Background}

Most approaches to traffic signal control rely on computer software for {\em microsimulation}, which simulates traffic at the level of individual vehicles, computing the position, velocity, emissions data and other for every vehicle at each time step.  A {\em route} is a sequence of roads used by vehicles to traverse the network.

\begin{figure}[!tb]
  \centering
  \includegraphics[width=0.9\columnwidth]{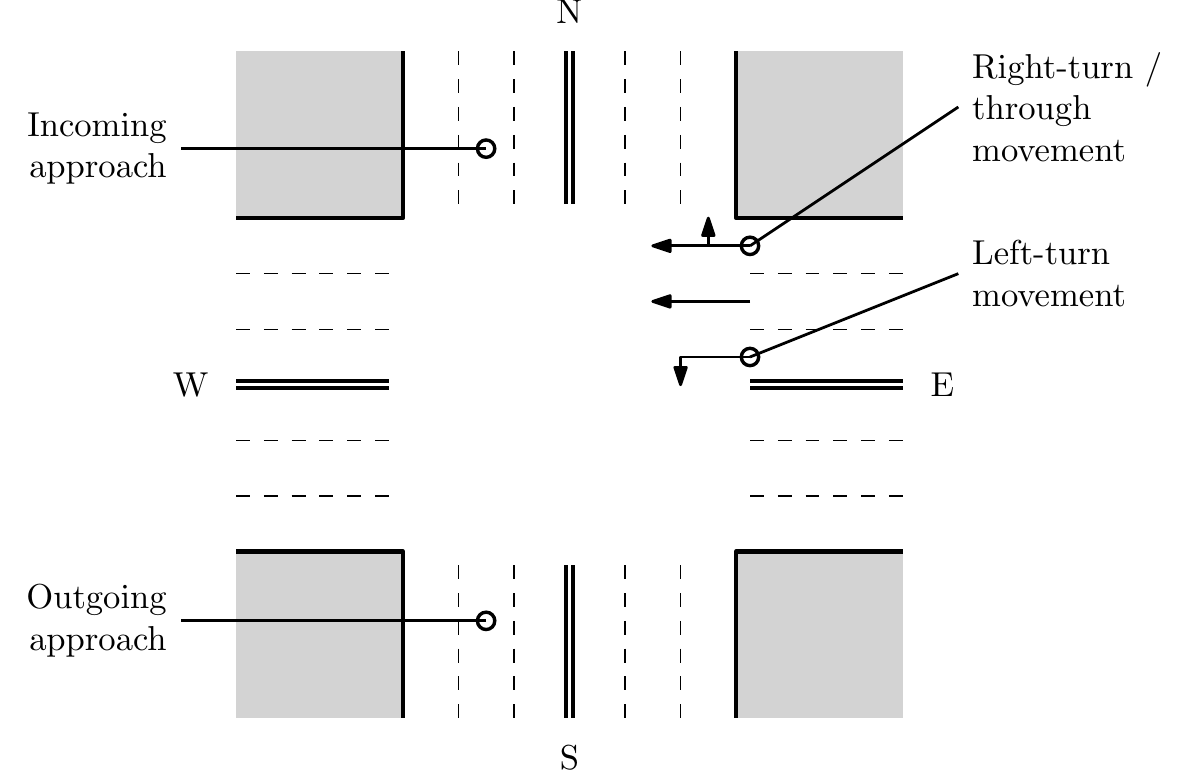}
  \caption{Intersection with four incoming approaches, each composed of three lanes.}
  \label{fig:terminology}
\end{figure}

A traffic infrastructure can be represented as a network/graph, where {\em roads} and {\em junctions} correspond to the edges and nodes, respectively. A road connects two nodes of the network and has a given number of {\em lanes}, a maximum speed and a length. Traffic light controllers are typically installed at road junctions. An intersection is a common type of junction in which roads cross each other. Figure~\ref{fig:terminology} illustrates a typical intersection with four incoming and four outgoing {\em approaches}, each approach composed of three lanes.

A {\em signal movement} refers to the transit of vehicles from an incoming approach to an outgoing approach. A signal movement can generally be sub-categorized as a {\em left-turn}, {\em through} or {\em right-turn} movement. For example, in the intersection of Fig.~\ref{fig:terminology}, East-South corresponds to a left-turn movement, while East-West corresponds to a through movement. Both are examples of signal movements. A green signal indicates that the respective movement is allowed, whereas a red signal indicates that the movement is prohibited.

A {\em signal phase} is a combination of non-conflicting signal movements, i.e. the signal movements that can be set to green at the same time. In the intersection of Fig.~\ref{fig:terminology}, the triplet (North through, South through, South right-turn) is a valid signal phase. In contrast, (North through, South left-turn) is not. A {\em signal plan} for a single intersection is a sequence of signal phases and their respective durations. Usually, the time to cycle through all phases, known as {\em cycle length}, is fixed. Therefore, the phase splits---i.e., the amount of time allocated for each signal phase---are normally defined as a ratio of the cycle length. A yellow signal is set as a transition from a green to a red signal.

\subsection{Reinforcement Learning} 

As mentioned in the introduction, reinforcement learning considers an agent whose interaction with the environment can be described as a {\em Markov decision process} (MDP). An MDP is a tuple $(\S, \A, \{\mathcal{P}_a, a \in \A \}, r, \gamma)$, where $\S$ is a set of {\em states} and $\A$ is a set of {\em actions}. At each step $t$, the agent observes the state $S_t\in\S$ of the environment and selects an action $A_t\in\A$. The environment then transitions to a state $S_{t+1}$, where
\begin{equation}
\PP{S_{t+1}=s'\mid S_t=s,A_t=a}=[\P_a]_{ss'}.
\end{equation}
The matrix $\P_a, a\in\A,$ encodes the {\em transition probabilities} associated with action $a$. Upon executing an action $a$ in state $s$, the agent receives a (possibly random) reward with expectation given by $r(s,a)$. The goal of the agent is to select its actions so as to maximize the {\em expected total discounted reward} (TDR),
\begin{equation}
\mathrm{TDR}=\EE{\sum_{t=0}^\infty\gamma^tR_t},
\end{equation}
where $R_t$ is the random reward received at time step $t$ (with $\EE{R_t}=r(S_t,A_t)$) and the scalar $\gamma$ is a {\em discount factor}. The long-term {\em value} of an action $a$ in a state $s$ is captured by the {optimal $Q$-value}, $Q^*(s,a)$, which can be computed using, for example, the {\em $Q$-learning algorithm} \cite{watkins_christopher_learning_1989}. The $Q$-learning algorithm estimates the optimal $Q$-values as the agent interacts with the environment: given a transition $(s,a,r,s')$ experienced by the agent, $Q$-learning performs the update
\begin{equation}\label{eq:q-learning}
\hat{Q}(s,a)\leftarrow \hat{Q}(s,a)+\alpha\big(r+\gamma\max_{a'\in\A}\hat{Q}(s',a')-\hat{Q}(s,a)\big),	
\end{equation}
where $\alpha$ is a step size. Upon computing $Q^*$, the agent can act optimally by selecting, in each state $s$, the optimal action at $s$, given by $\pi^*(s)=\argmax_aQ^*(s,a)$. The mapping $\pi^*:\S\to\A$, mapping each state $s$ to the corresponding optimal action $\pi^*(s)$, is known as the {\em optimal policy} for the MDP.

\textit{Deep Q-network} \cite{mnih_2013} is a well known RL method that approximates the Q-values with a neural network $\hat{Q}(s,a;\theta)$, where $\theta$ denotes the parameters of the model. At each step, the agent adds a transition $(s,a,r,s')$ to a replay memory buffer, from which batches of transitions are sampled in order to optimize the parameters of the model such that the following loss in minimized:

\begin{equation}
    \label{eqn:DQN_update_eq}
    \mathcal{L}(\theta) = \left(r + \gamma\max_{a'\in\A}\hat{Q}(s',a';\theta^{-})  - \hat{Q}(s,a;\theta)\right)^2.
\end{equation}

\noindent The gradient of the loss is backpropagated only into the \textit{behaviour network}, $\hat{Q}(s,a;\theta)$, which is used to select actions. The term $\theta^{-}$ represents the parameters of the \textit{target network}, a periodic copy of the behaviour network.


\section{Related Work}%
\label{Sec:Related-work}

Several works have explored the use of RL in traffic light control, most of which rely on estimating the $Q$-function or an approximation thereof  \cite{abdulhai_reinforcement_2003,wiering_multi-agent_2000}. In their simplest form, such RL approaches consider that each intersection is controlled by a single agent that ignores the existence of other agents in neighboring intersections. More sophisticated approaches consider the existence of multiple agents and leverage the network structure to address the interaction between the different agents \cite{prabuchandran14itsc,liu17itsc}. With the advent of deep learning, several of the approaches above have been extended to accommodate deep neural networks as the underlying representation of the problem. For example, Genders and Razavi \cite{Genders16} propose a DQN control agent that combines a deep convolutional networks with $Q$-learning. The work controls the traffic lights at a single intersection, and essentially extends previous work to accommodate the deep learning model. 


In terms of experimental methodology, there are several issues that make the comparison of different approaches challenging. First, different works adopt different evaluation metrics and baseline policies. In one work, the performance metrics used are the average travel time per car and the average wait time per car; the proposed approach is compared against a fixed timed plan \cite{thorpe_vehicle_1997}.  In a different work, the metrics adopted are the average waiting time and number of refused cars (a saturation condition of the used simulator), and the proposed approach is compared against both a random policy and a fixed-time policy \cite{wiering_multi-agent_2000}. In yet another work, the metrics used are the ratio of the average delay, and the proposed approach is compared against a pre-timed plan \cite{abdulhai_reinforcement_2003}. The lack of a clear understanding of the dependence of the different metrics on the number of intersections is another issue that renders comparisons difficult. Finally, while average metrics of the variables of interest are provided, most works offer no measures of significance regarding the reported performance.

Recently, several works sought to address some of the issues previously discussed \cite{genders_open-source_2019}. Researchers have put forth several recommendations/good practices when exploring the use of RL in the context of ATSC: (i) provide all hyper-parameters and number of trial experiments; (ii) report aggregated results with deviation metrics (averages and standard deviations, not maximum returns); (iii) implement proper experimental procedures (average together many trials using different random seeds for each) \cite{islam_reproducibility_2017}.
We extend those works by providing both the preliminary steps necessary to simulate, develop, train and evaluate RL-based experiments for ATSC on real world scenarios, as well as, insights which can be extracted by our methodology in this domain.


\section{Methodology}%
\label{Sec:Methodology}

\begin{figure*}[t]
\centering
\includegraphics[width=0.75\textwidth]{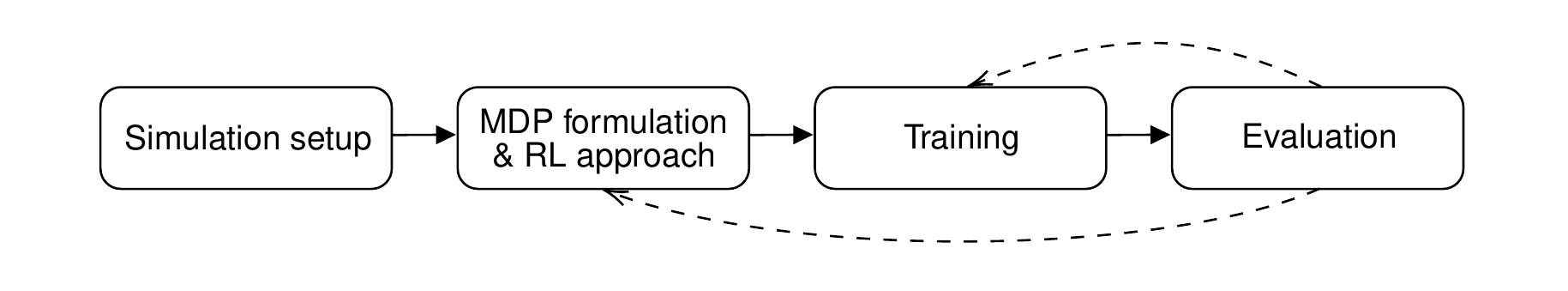}
\caption{Diagram illustrating the proposed methodology, composed of four stages. Solid arrows denote the main development flow, whereas dashed arrows denote the iterative process of model tuning.}
\label{fig:methodology}
\end{figure*}

We propose a four-stage methodology to be used in the development of RL-based traffic signal controllers. Figure ~\ref{fig:methodology} illustrates the proposed methodology, comprising four phases: simulation setup, MDP formulation and selection of the RL learning method, train and evaluation.

\subsection{Simulation setup}

The first stage of the proposed methodology is the {\em simulation setup} phase. RL controllers must be trained by resorting to (micro-)simulators that are able to provide a realistic response to the agent's actions during the learning process. The main objective of the simulation setup stage is to prepare all the simulations needed to carry out such training. It includes gathering simulation-related data such as the topological data of the roads' network and vehicles demand/routes data, as well as setting up the traffic simulator. 

In any case, for the purpose of our methodology, it is important that during this stage two key components are configured and defined: (i) the topology of the roads network; and (ii) the traffic demands and routes.

\subsubsection{Roads network topology.}

Networks can be either synthetic or extracted from real world locations. Available open source services, such as {\sc OpenStreetMap} \cite{OpenStreetMap}, allow segments of cities' districts to be exported and, during the simulation setup step, such information can be prepared and fed to the simulator, thus opening up the possibility of simulating a rich set of networks relevant to real-world traffic signal control. 

In our own implementation, we use geospatial data from {\sc OpenStreeMaps} to build the configuration files to be used by the simulator, in our case, the SUMO micro-simulator \cite{krajzewicz12ijasm}. The following steps are required: (i) extract the region of interest from {\sc OpenStreetMap} and open the resulting file with the {\sc JOSM}\footnote{\url{https://josm.openstreetmap.de/}} editor, an extensible editor for {\sc OpenStreetMap} files, in order to fine-tune the network; (ii) convert the (edited) {\sc OpenStreetMap} file into the SUMO network format using the \texttt{netconvert}\footnote{\url{https://sumo.dlr.de/docs/netconvert.html}} tool; and (iii) open the resulting SUMO network file with the \texttt{netedit}\footnote{\url{https://sumo.dlr.de/docs/netedit.html}} tool, a graphical network editor for SUMO, in order to ensure that all intersections are properly setup, namely check whether all traffic phases and links (connections between lanes) are correct.

\subsubsection{Traffic demands and routes.}
 Traffic demands and routes can be either synthetic \cite{wei_2018} or derived from real-world data using origin-destination matrices \cite{Aslani17} or induction loops counts \cite{TLC_robustness}. Regarding synthetic demands, simple (constant demands) to complex (variable demands) scenarios can be created by specifying the probabilites of vehicles' insertion through time. With respect to the generation of a synthetic set of routes, the \texttt{duarouter}\footnote{https://sumo.dlr.de/docs/duarouter.html} tool can be used. Afterwards, a probability can be assigned to each unique route by weighting it accordingly to a pre-determined criteria, such as, invertionally proportional to the number of turns contained in the respective route.


\subsection{MDP formulation and RL approach}
The second stage in our methodology is the description of the traffic control problem as a Markov decision problem (or a multiagent version thereof). As previously discussed, an MDP comprises 5 elements: the set of {\em states}, the set of {\em actions}, the {\em transition probabilities}, the {\em reward}, and the {\em discount factor}.

Most works which apply RL to the ATSC domain do not specify the transition probabilities since they are not strictly required -- additionally explicitly modeling the traffic dynamics as a result of changes in traffic light control is unfeasible in most cases. The remaining four components must still be specified: the state space (i.e., the information upon which the agent will base its decisions), the action space (i.e., how the agent is able to influence the environment through the choice of its actions) and the reward function (which implicitly encodes the goal of the agent). The literature is very diverse with respect to the adopted problem formulation; it is important to stress that the performance of the resulting controller will depend critically on the choices made at this stage. 

Alongside the MDP formulation, it is also necessary to select the desired RL method to be used as a learning component for the traffic signal controller, a choice that is not independent of MDP formulation. The literature is also very diverse with respect to the algorithm choice, even though the majority of the works focus their attention on the study of value-based methods (i.e., methods that seek to estimate/approximate the $Q$-function or a surrogate thereof). 

The work of El-Tantawy et al. \cite{el-tantawy_design_2014} provides a comparison between some common state-space representations, reward functions and action space definitions, using a real-world network topology. Wei et al.\ \cite{wei_survey_2020} provide a comprehensive list of commonly used MDP formulations in the context of ATSC, as well as RL methods used in the context of traffic signal control.


\subsection{Training}

The \textit{training} procedure of RL-based traffic signal controllers should follow the same guidelines used in the field of RL. For example, a proper balance between exploration and exploitation should be ensured, making sure that the agent is able to experience a wide range of different situations. In adherence to the good practices previously discussed, it is important to run multiple instances of the training process, using different seeds, in order to correctly assess the learning ability of the proposed RL method.

In the context of ATSC, particular attention must be paid to ensure that the simulations are properly running. {\em Grid-locks} should be avoided or properly processed, for example by restarting the simulation, adjusting the vehicles' arrivals, or teleporting vehicles.%
\footnote{A gridlock occurs when a queue from one bottleneck creates a new bottleneck somewhere else, and so on in a vicious cycle that completely stalls the vehicles' circulation \cite{daganzo07trb}.}

%

Finally, it is important to monitor performance metrics such as losses, rewards, the number of vehicles in the simulation and the vehicles' velocity throughout the training in order to gain a better insight into the learning process.

The outcome of the training process consists of a set of policies (each one resulting from a different training run), that need to be properly evaluated. The next and final step of the proposed methodology addresses how this can be accomplished.


\subsection{Evaluation}

In the context of traffic light control, several performance metrics have been proposed: travel time, waiting time, number of stops, queue length, throughput, as well as gas emissions and fuel consumption. From all these metrics, the minimization of the travel time is usually the main goal in the development of ATSCs, therefore, it is arguably the most important metric to report. Since algorithms are usually unable to directly optimize the travel time, it is useful to report additional metrics that are more closely related with the formulated agent’s objective (the reward function, in the case of RL agents), such as the queue length or waiting time.

It is important to run some baseline algorithms, such as the Webster or Max-Pressure methods, under the same scenario. These runs are of extreme importance since they allow to compare the performance of the RL controller(s) against well-established, commonly used traffic engineering methods.

\subsubsection{Performance estimation.}

In order to adequately compare the different approaches, it is important to assess the performance of the alternative proposals with a set of accordingly seeded simulations in order to rule out any influence of the simulation seeds in the results. For the baseline algorithms, this can be achieved by simply running multiple evaluation simulations. With respect to the RL agents, each of the policies that resulted from the training stage should be evaluated with a set of evaluation rollouts and, if posteriorly needed, the results aggregated per policy.

\subsubsection{Performance analysis \& comparison.}

The simplest and most straightforward way to present and compare the performances of the different methods is through point estimation, for example by reporting the mean and standard deviation of the travel times observed for a set of evaluation rollouts. While this is commonly used in the ATSC domain, sometimes it exists a big overlap in the reported performance metrics between different methods, thus, it is important to understand whether the observed differences are statistically significant or not.

A better insight into the performance of the methods can be achieved by plotting the distributions of the travel time means for each of the methods, however, in order to draw conclusions from the observed performances it is proposed the use of statistical hypothesis testing \cite{ross2004introduction}. Specifically, we are interested in trying to understand whether two or more population means (the performance metrics of the different methods) are equal. It is highly likely that the previously hypothesis is rejected, therefore, \textit{post hoc} comparisons need to be performed in order to understand between which mean pairs exists a statistically significant difference. In order to perform such evaluation, it is proposed the use of the (one-way) ANOVA test, followed by the Tukey Honestly Significant Difference (HSD) test for \textit{post hoc} comparisons. Unfortunately, the previous tests make some assumptions related with the data distributions that not always hold. Therefore, it is worth checking whether the assumptions are met, and if not, to switch the aforementioned tests with their respective non-parametric versions.

As a complementary analysis, it might be interesting to plot histograms for the performance metrics as most of the times, mean values may not be well representative of the underlying distributions.





\section{Experiments}%
\label{Sec:Experiments}

We now illustrate the application of the previously described methodology by developing a DQN-based ATSC for two real-world scenarios in Lisbon, Portugal.


\subsection{Pre-processing}

\begin{figure}[!tb]
\centering
\includegraphics[width=0.42\textwidth]{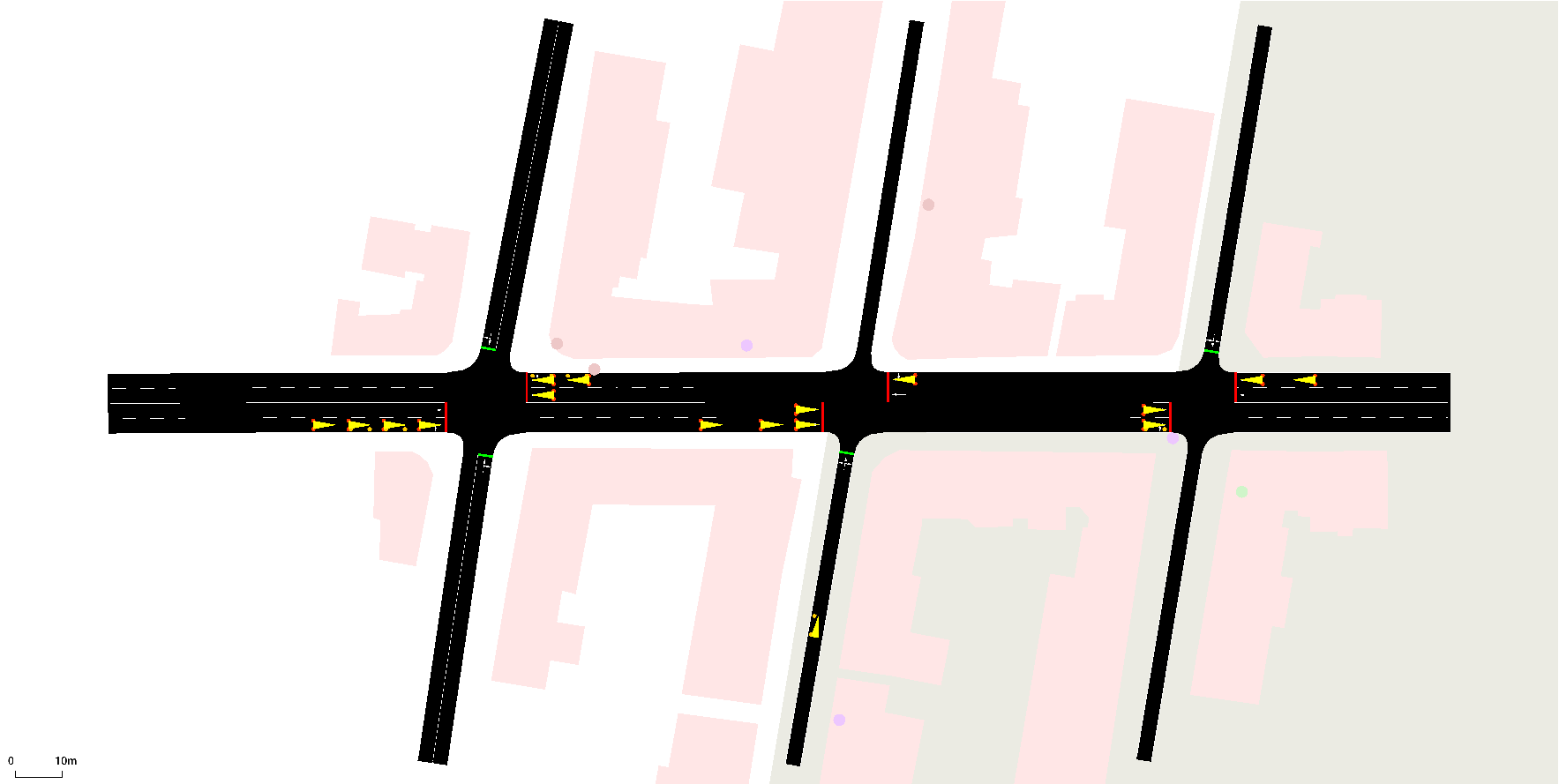}
\caption{SUMO screenshot of the considered intersections, near Marqu\^{e}s de Pombal square, in Lisbon. Scenario 1 comprises only the left-most intersection whereas scenario 2 is composed of all three pictured intersections. Phase 1 allows the movement of vehicles in the vertical direction whereas phase 2 allows the movement in the horizontal direction. }
\label{fig:scenarios}
\end{figure}

For the purposes of this work, two real-world scenarios in the Lisbon metropolitan area are considered: scenario~1  consists of a single intersection, whereas the second scenario consists of a set of three consecutive intersections. The two scenarios were extracted from {\sc OpenStreetMap} by following the previously described steps. The JOSM editor was used to further crop, rotate and resize the area of interest. The resulting SUMO environment is shown in Figure \ref{fig:scenarios}.  The considered demands are time-constant and proportional to the number of lanes. The routes are weighted using the previously described synthetic procedure.


\subsection{MDP formulation and RL approach}

We adopt a simple approach where an RL agent controls a single intersection, ignoring the existence of other agents in neighboring intersections \cite{abdulhai_reinforcement_2003}. In other words, each individual agent is modeled using an MDP that considers only the traffic information in the intersection controlled by that agent.

In each intersection, the signal cycle length is fixed to 60~seconds, and the yellow time to 6~seconds. At the beginning of each cycle, the controller is able to pick the signal plan to be executed throughout the next cycle, from a set of predefined signal plans. In our case, all intersections are composed of two phases. More precisely, the \textit{action space} consists of a discrete set of 7 signal plans \{0: (30\%, 70\%), 1: (37\%, 63\%), 2: (43\%, 57\%), 3: (50\%, 50\%), 4: (57\%, 43\%), 5: (63\%, 37\%), 6: (70\%, 30\%) \}, where the first and second elements of each tuple correspond, respectively, to the phase split of phase 1 and phase 2. With this action space definition, adjacent traffic controllers can be easily synchronized and a minimum green time is guaranteed for all phases, easily ensuring that all safety standards are met.

The state $s$, at cycle $c$ is represented by the tuple $(w_1, w_2)$, where $w_p$ is the cumulative number of vehicles waiting, or navigating at low speeds, in phase $p$, and can be computed according to:

\begin{align}\label{eqn:waiting_time}
        w_p &= \frac{1}{K}\sum_{k=0}^{K-1}\sum_{l\in L_{p}}\sum_{v\in V^k_l}stopped(v, k),
\end{align}

\noindent where $K$ is the cycle length in seconds (fixed as $K=60$), $L_p$ is the set of all inbound lanes to phase $p$, and $V^k_l$ is the set of vehicles on $l$ lane at time $k$. A vehicle is said to be stopped when it has a very low speed:

\begin{equation}\label{eqn:stopped}
        stopped(v, k) =
          \begin{cases}
           1       & \quad speed(v, k) < \text{threshold}\\
           0       & \quad \text{otherwise}
          \end{cases}
\end{equation}

\noindent where the used threshold corresponds to 10\% of the maximum velocity and all vehicles' speeds are normalized as to belong to the interval $[0,1]$.

Finally, we use an action-independent reward function defined, for a state $s=(w_1,w_2)$ in intersection $i$ and cycle $c$ with phases $P$, as:
\begin{equation}\label{eqn:reward}
r = -\sum_{p\in P} w_p
\end{equation}

The reward in Eq. \eqref{eqn:reward} consists of the (negative) sum of the total amount of seconds the vehicles have been stopped during the cycle. We use $\gamma=0.95$. 



\subsection{Training}

\begin{figure}[!tb]
\centering
\includegraphics[width=0.35\textwidth]{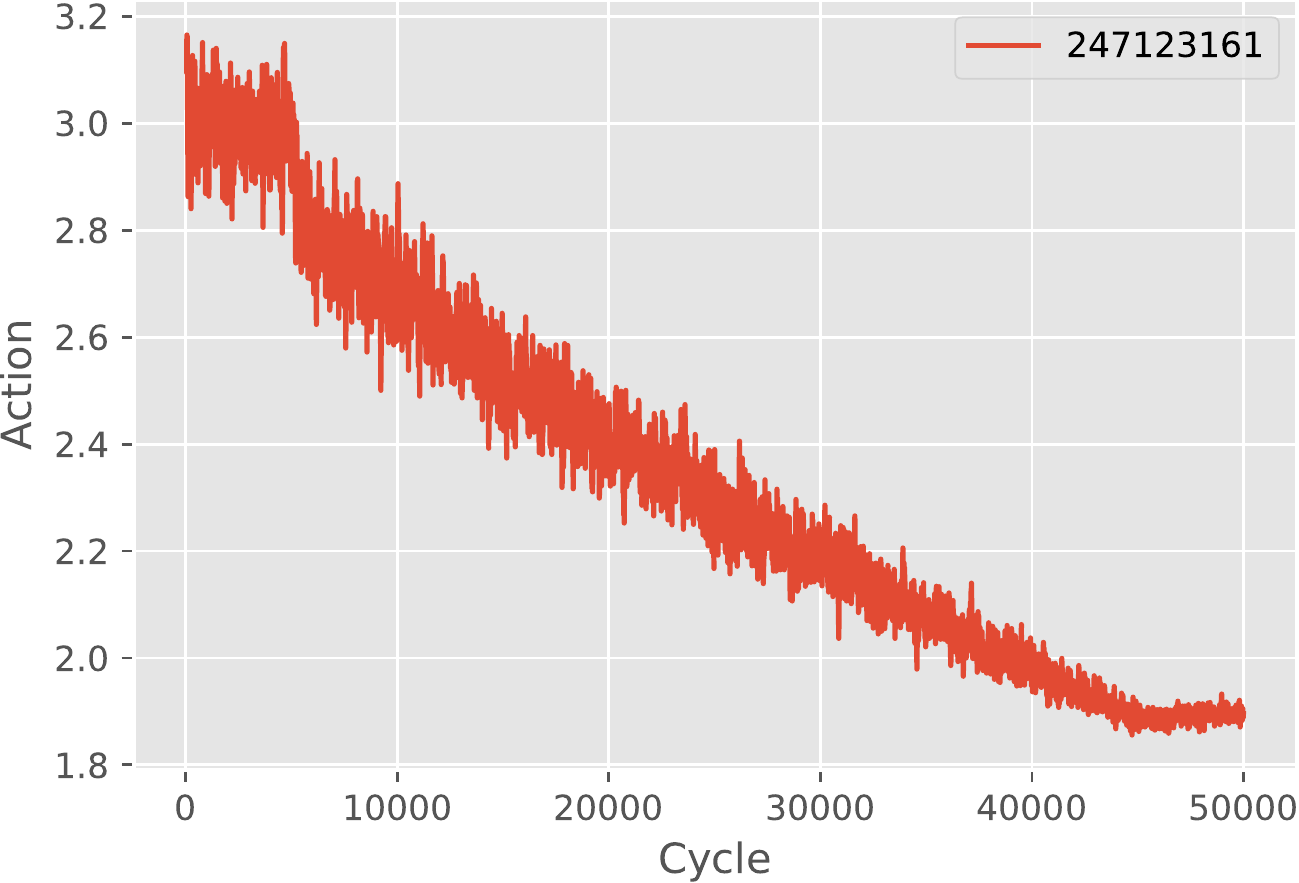}
\caption{(Scenario 1) Mean actions.}
\label{fig:intersection_train_actions}
\end{figure}

\begin{figure}[!tb]
\centering
\includegraphics[width=0.35\textwidth]{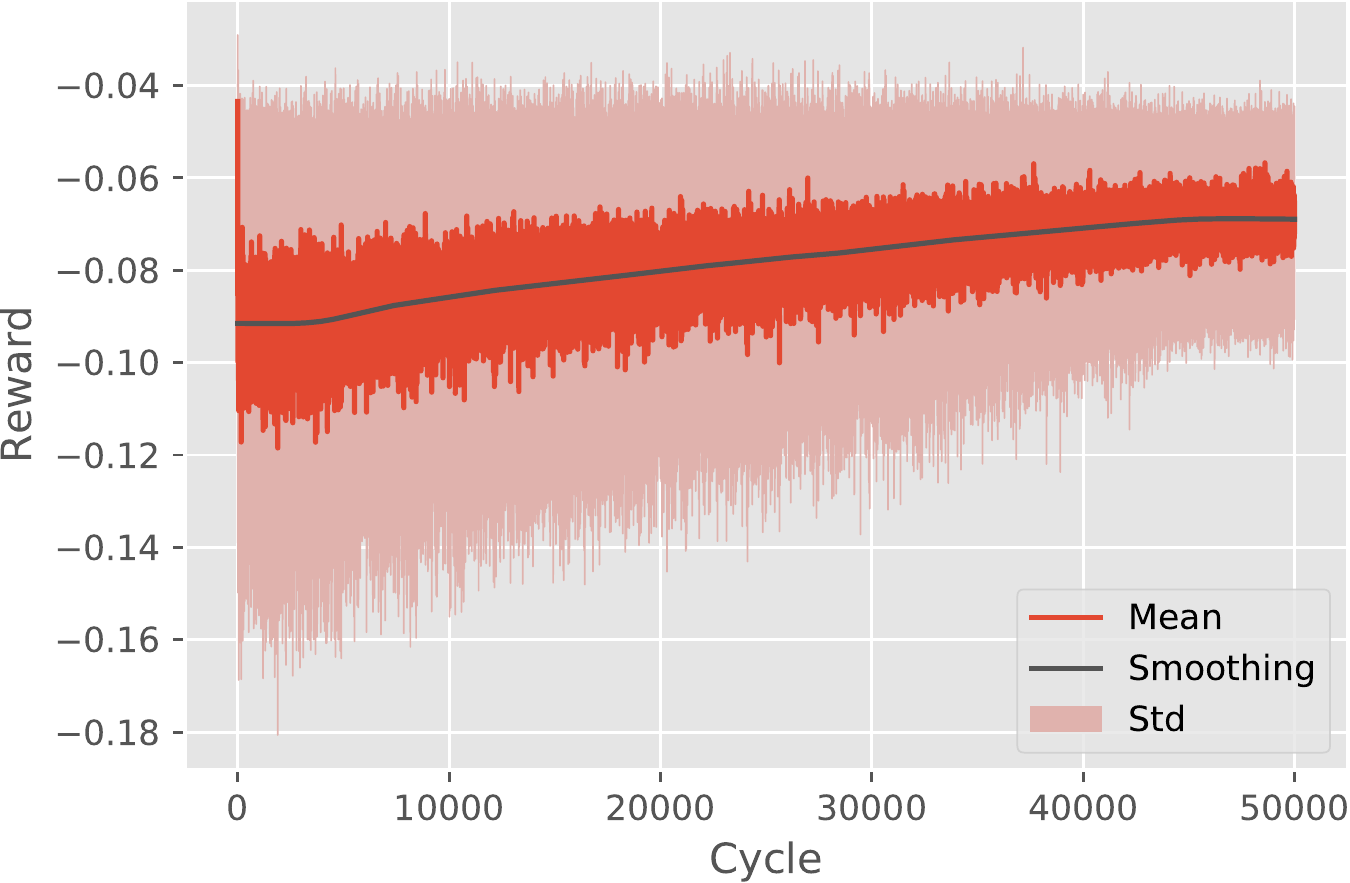}
\caption{(Scenario 1) Instantaneous rewards.}
\label{fig:intersection_train_rewards}
\end{figure}

We ran 30 independent training runs. Figures \ref{fig:intersection_train_actions} and \ref{fig:intersection_train_rewards} display, respectively, the mean actions selected during training and the observed instantaneous rewards. As it can be seen, the agent's actions converge towards lower-indexed signal plans, which can be justified by the intersection's structure (Fig. \ref{fig:scenarios}): the horizontal direction serves more vehicles, therefore the agent converges towards lower-indexed actions which allocate a longer period to this phase. As the actions converge it is noticeable an increase in the observed instantaneous rewards.

The outcome of the training stage consists of 30 policies that need now to be properly assessed.

\subsection{Evaluation}

Finally, the performance of the resulting agents is evaluated using a set of performance metrics: travel time, waiting time and vehicles' speed. Tables~\ref{table:results_intersection} and \ref{table:results_grid} display the evaluation metrics for different traffic signal controllers. The actuated controller 
dynamically extends the current phase, up to a maximum value, if a continuous stream of incoming vehicles is detected.

With respect to scenario 1 (Tab. \ref{table:results_intersection}), the results show that the RL controller is able to achieve the highest average speed, outperforming all the other controllers. With respect to the travel time metric, the RL agent is able to outperform the Webster method and equal the performance of both the best static controller and the actuated method. However, the agent is unable to outperform the Max-Pressure controller, but it is important to notice that this method exhibits a higher degree of flexibility in comparison to the RL agent since it´s cycle length is dynamic.

Regarding the second scenario (Tab. \ref{table:results_grid}), the RL agents are again able to achieve the highest average speed. Regarding the travel time, it can be seen that the RL agents are able to outperform the actuated method as well as the Max-Pressure controller, despite the fact that these methods are able to achieve a significantly lower waiting time. This happens due to miscoordination between the adjacent intersections for both the actuated and Max-Pressure methods (due to their acyclic behaviour).

\begin{table}[t]
\centering
\resizebox{.95\columnwidth}{!}{
    \begin{tabular}{ | c | c | c | c |}
    \hline
    \textbf{Method}	& \textbf{Speed} & \textbf{Waiting time} & \textbf{Travel time}\\ \hline \hline
    Static  & (6.7, 3.5) & (8.1, 9.9) & (25.0, 12.5) \\ \hline
    Webster & (6.6, 3.4) & (8.2, 9.7) & (25.4, 12.4) \\ \hline
    Max-pressure  & (6.5, 2.8) & \textbf{(5.7, 6.5)} & \textbf{(23.4, 9.0)} \\ \hline
    Actuated  & (6.7, 3.4) & (7.8, 10.2) & (24.9, 12.8) \\ \hline
    RL controller  & \textbf{(6.8, 3.5)} & (8.0, 10.1) & (25.0, 12.6) \\ \hline
    \end{tabular}}
\caption{(Scenario 1) Evaluation metrics aggregated per vehicle's trip (averaged over 30 simulations). The first tuple position encodes the mean value; the second tuple position encodes the standard deviation.}
\label{table:results_intersection}
\end{table}

\begin{table}[t]
\centering
\resizebox{.95\columnwidth}{!}{
    \begin{tabular}{ | c | c | c | c |}
    \hline
    \textbf{Method}	& \textbf{Speed} & \textbf{Waiting time} & \textbf{Travel time}\\ \hline \hline
    Webster & (6.3, 2.7) & (12.0, 12.0) & \textbf{(38.6, 14.7)} \\ \hline
    Max-pressure  & (5.8, 2.2) & \textbf{(9.0, 7.0)} & (42.3, 18.6) \\ \hline
    Actuated  & (5.6, 2.5) & (11.6, 10.5) & (45.5, 22.0) \\ \hline
    RL controller  & \textbf{(6.3, 2.8)} & (12.2, 10.9) & (39.1, 15.5) \\ \hline
    \end{tabular}}
\caption{(Scenario 2) Evaluation metrics aggregated per vehicle's trip (averaged over 30 simulations). The first tuple position encodes the mean value; the second tuple position encodes the standard deviation.}
\label{table:results_grid}
\end{table}

\begin{figure}[h]
\centering
\includegraphics[width=0.35\textwidth]{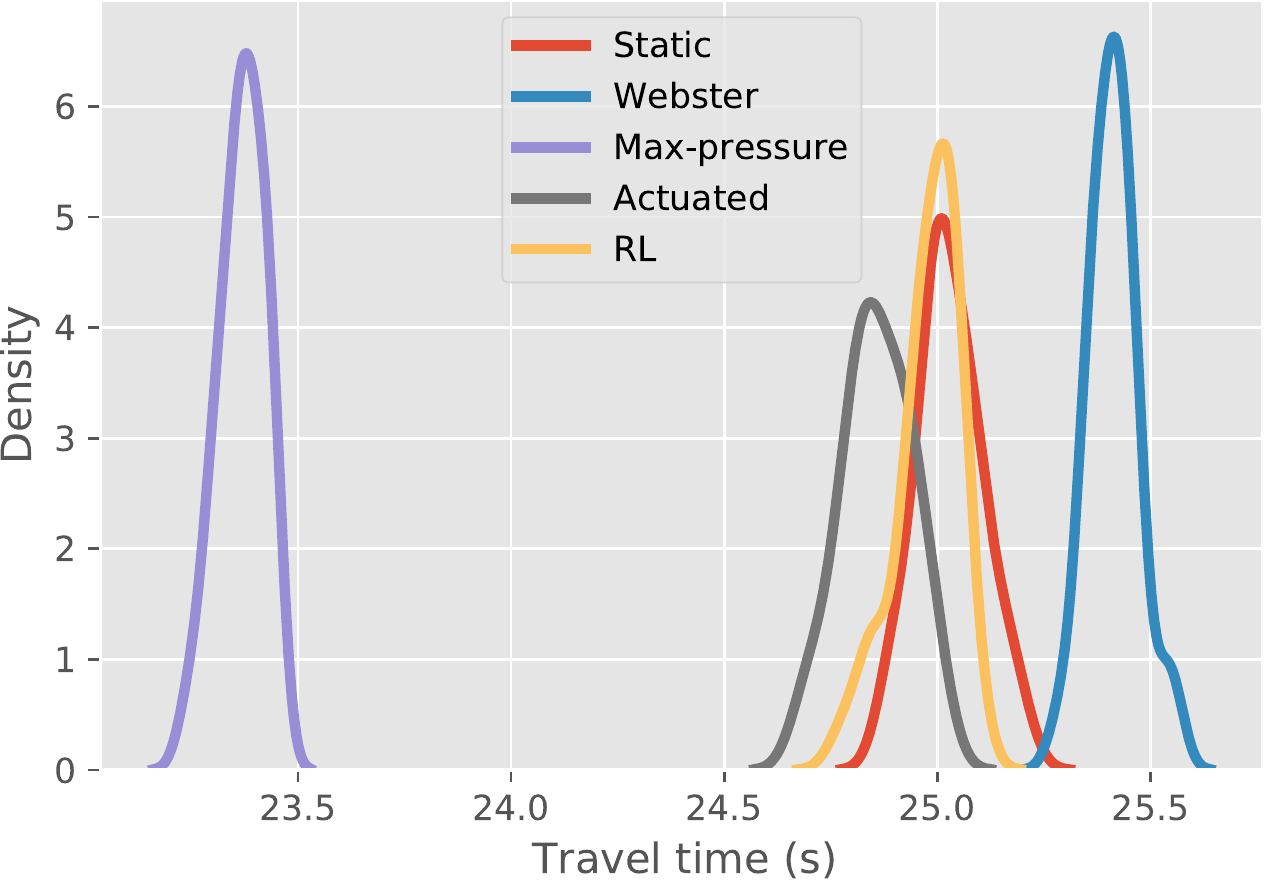}
\caption{(Scenario 1) Kernel density estimation of the travel time means, computed using 30 samples for each of the methods.}
\label{fig:travel_time_kde}
\end{figure}

Fig. \ref{fig:travel_time_kde} displays the distribution of the travel time means for scenario 1. As it can be seen, there is a significant overlap in performance between the different methods. Furthermore, despite the fact that the ANOVA test yields a p-value of $\approx 0$ (meaning that, indeed, the mean performance of all methods is not the same), the Tukey HSD pairwise tests show no significant difference in mean performance between some methods. Namely, between the actuated, static and RL controllers, the confidence interval on the means' difference either includes zero, or its bounds are close to it.

\begin{figure}[h]
\centering
\includegraphics[width=0.35\textwidth]{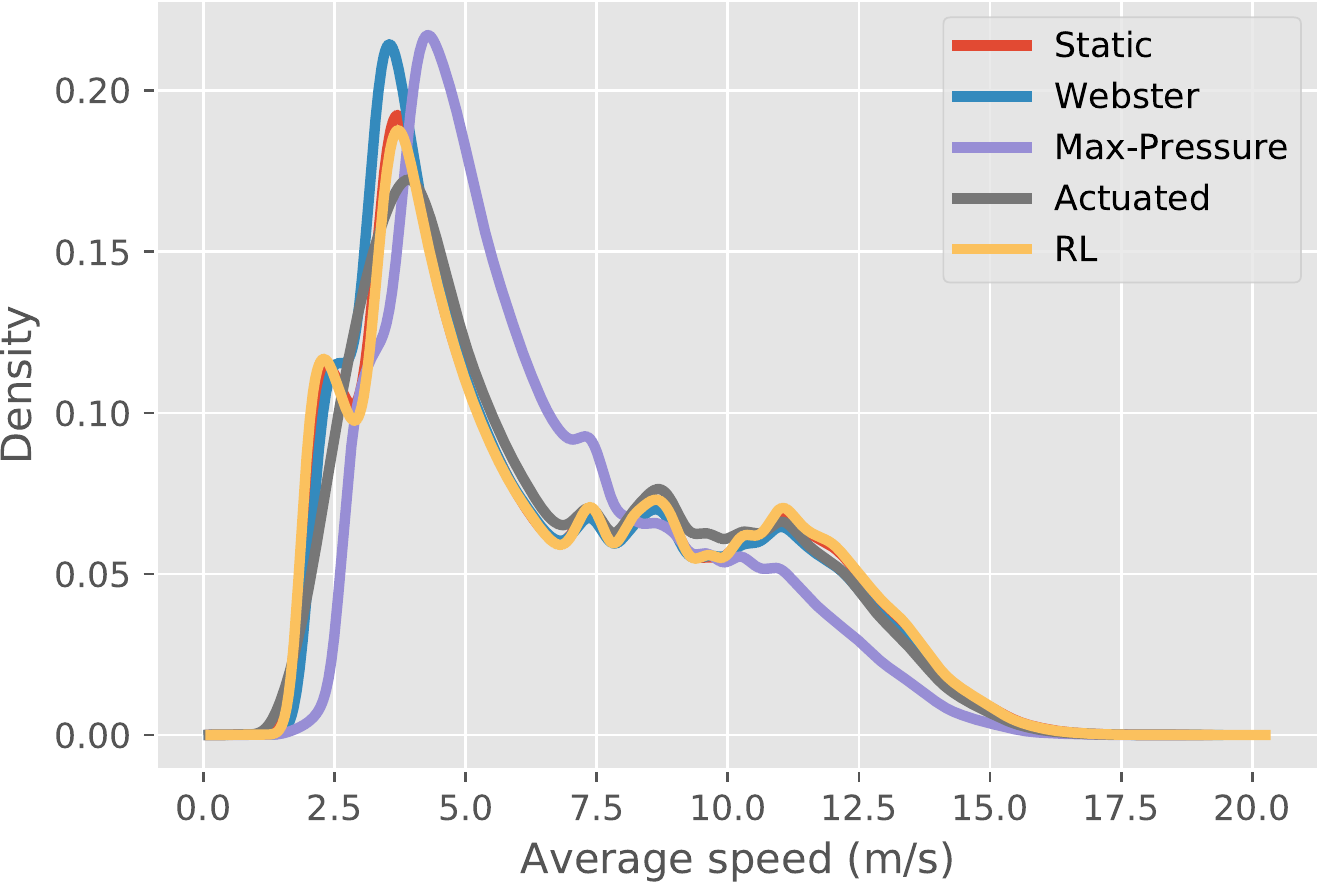}
\caption{(Scenario 1) Kernel density estimation of the vehicles' speeds, computed using 30 samples for each of the methods.}
\label{fig:speed_kernel}
\end{figure}

Finally, Fig. \ref{fig:speed_kernel} displays the distribution of the vehicles' speed per trip. Interesting to notice the multimodal shape of the underlying distributions, as well as the slight differences between the different methods.

\section{Conclusions}
In this paper, we proposed a methodology for the development of RL-based adaptive traffic signal controllers. The proposed methodology comprises 4 steps, all of which necessary to develop, deploy and evaluate an ATSC --- simulation setup, problem formulation, training and evaluation. We illustrated the proposed methodology by developing a deep RL-based ATSC that achieves performance on par with established methods from the transportation engineering field. Despite the fact that the flexibility of the agent is constrained in order to met safety standards, the presented results glimpse at the potential of RL-based controllers to contribute to improve traffic congestion and highlight the need for coordination between adjacent intersections. Finally, we note that the advantages of RL-methods are more apparent in scenarios comprising bigger traffic networks and more variable traffic patterns, something that could be considered in future work.



\section{ Acknowledgments}
This work was partially supported by national funds through the Portuguese Funda\c{c}\~{a}o para a Ci\^{e}ncia e a Tecnologia under project UIDB/50021/2020 (INESC-ID multi-annual funding) and the project ILU, with reference DSAIPA/DS/0111/2018. This research was also partially supported by TAILOR, a project funded by EU Horizon 2020 research and innovation programme under GA No 952215.



\end{document}